\documentclass{iopjournal}

\usepackage{upgreek}

\begin{document}


\title{Why the better conductor surges in impedance:\\
a low-cost demonstration of the skin effect at line frequency}

\author{Florian Platten$^{1,*}$ and Helmut Wenz$^{1}$}

\affil{$^1$Faculty of Mathematics and Natural Sciences, Heinrich Heine University D\"{u}sseldorf, D\"{u}sseldorf, Germany}

\affil{$^*$Author to whom any correspondence should be addressed.}

\email{florian.platten@hhu.de}


\begin{abstract}
When alternating current flows through a conductor, induced internal eddy currents suppress current flow in the core and restrict it to a thin surface layer. 
Although this effect, known as the skin effect, is covered in virtually every undergraduate electrodynamics course, it is rarely explored experimentally without specialized and expensive equipment, and is sometimes even considered a mere high-frequency phenomenon. 
Here, we present a readily accessible, quantitative demonstration of the skin effect at line frequency using only standard laboratory supplies: an AC/DC power supply, basic multimeters, and laboratory support rods as cylindrical conductors. 
Based on a precise four-point Kelvin sensing scheme capable of resolving milliohm-range resistances, we exploit relative magnetic permeability rather than high frequency as the governing parameter to control skin depth.
For non-magnetic stainless steel, line frequency represents an electrodynamically low frequency, resulting in identical AC and DC behaviour; whereas for ferromagnetic structural steel, high permeability compresses the skin depth to the millimeter scale, causing a sharp surge in AC impedance. 
This produces a counterintuitive outcome where the statically superior conductor suffers a marked relative impedance increase, whereas the poorer conductor remains unaffected, challenging common student expectations and catalyzing cognitive activation.
Depending on the instructional setting, the approach can be implemented either as a quick lecture demonstration or as a quantitative experiment within an undergraduate laboratory course.

\end{abstract}

\section{Introduction}

In physics and electrical engineering education, 
the transition from steady-state direct current (DC) to 
time-varying electrodynamics, where induction drives internal eddy currents, constitutes a major conceptual threshold \cite{Galili,Guisasola,Guisasola2}. 
A central phenomenon in this regime is the skin effect \cite{Simonyi,Landau}: the tendency of an alternating electric current to distribute itself unevenly within a conductor, resulting in a higher current density near the surface that decays exponentially towards the core. This spatial redistribution effectively reduces the usable cross-sectional area. 
Consequently, the alternating-current (AC) resistance $R_{\mathrm{AC}}$ increases with angular frequency $\omega$ above its static direct-current baseline $R_{\mathrm{DC}}$, yielding: 
\begin{equation}
R_{\mathrm{AC}}(\omega) \ge R_{\mathrm{DC}}
\end{equation}
with $R_{\mathrm{AC}}(0) = R_{\mathrm{DC}}$.

Understanding this spatial current redistribution is essential not only for foundational electrodynamics, but also for practical engineering applications \cite{Fink}. Most prominently, it dictates the design of high-frequency waveguides. In these hollow metallic pipes, electromagnetic wave propagation is driven solely by thin surface currents along the inner walls, making core material redundant. In power engineering, it dictates high-power busbar design, where heavy-duty conductors use flat or multi-leaf geometries to maximize surface area and minimize ohmic losses during high-current distribution. Finally, in manufacturing, it enables induction surface hardening, a process using alternating magnetic fields to heat and harden only the outer skin of a component while leaving its ductile core intact.

Despite its technical importance, a survey of contemporary instructional literature reveals a striking asymmetry in how eddy currents and the skin effect are demonstrated in the classroom. Qualitative demonstrations of induced magnetic forces are ubiquitous. Most prominently, the Waltenhofen pendulum illustrates Lenz’s law via magnetic braking: as a conductive plate swings into a magnetic field, induced eddy currents generate an opposing Lorentz force that abruptly damps its motion. Historically built with massive U-core electromagnets \cite{Lueders}, such setups are today easily implemented using compact Neodymium magnets \cite{Henschke_Induction,PASCO_Magnet}. Yet, while visually striking, these mechanical demonstrations provide no access to the underlying alterations in electrical transport properties.

Where the skin effect itself is addressed experimentally, demonstrations remain largely qualitative, focusing on field attenuation or current displacement. 
To demonstrate magnetic shielding, classic setups insert thin aluminum sheets into high-frequency fields, generated by Tesla transformers \cite{3B_Tesla} or function-generator-driven coaxial coils \cite{UCB_SkinEffect}. Intercepting the field visibly extinguishes an indicator lamp on a receiver coil, demonstrating that shielding efficiency increases with frequency.
Alternatively, a cage-like conductor arrangement comprising a central rod surrounded by a ring of parallel outer wires, each equipped with inline indicator lamps, visually contrasts current transport modes \cite{Lueders}: direct DC application illuminates all lamps uniformly, whereas high-frequency AC restricts current strictly to the outer perimeter, leaving the central wire dark.
Further experimental demonstrations can be found in \cite{Pohl}.
While effective as conceptual visual aids, none of these demonstrations enable students to quantitatively measure the actual change in the ohmic resistance of the conductor.

To the best of our knowledge, quantitative $R_{\mathrm{AC}}$ experiments on standard laboratory equipment are virtually absent from introductory physics collections. 
Dedicated high-frequency setups are dominated by inductive reactance, making it challenging to extract the small  milliohm-scale resistance from the large reactive background.
Furthermore, they typically require costly instrumentation, like lock-in amplifiers or precision impedance analyzers, that is rarely available for large student cohorts and risks acting as a ``black box" that obscures the core electrodynamic mechanisms.

To bridge this gap, this paper presents an accessible and cost-effective experimental approach that, beyond a static DC baseline, operates exclusively at standard power-grid frequency ($50\,\mathrm{Hz}$). 
Rather than sweeping frequencies over several orders of magnitude with specialized high-frequency equipment, the relevant electrodynamic length scale, the skin depth $\delta$, is tailored via the magnetic permeability of the specimen. 
By contrasting ferromagnetic structural steel with non-magnetic (austenitic) stainless steel, we induce a counterintuitive effect: 
the conductor of lower DC resistivity displays a pronounced AC impedance surge at line frequency, whereas the higher-resistance sample remains essentially unaltered. 
High-current excitation from standard variable AC/DC power supplies amplifies both DC and AC voltage drops into easily measurable millivolt signals, while a four-point Kelvin scheme eliminates parasitic contact resistances. 

Following a brief theoretical framework (Section~\ref{sec:theory}), this paper presents the experimental setup (Section~\ref{sec:setup}), quantitative results (Section~\ref{sec:results}), and pedagogical implications (Section~\ref{sec:didactics}).

\section{Theoretical background}
\label{sec:theory}

This section outlines the electrodynamic foundation required to interpret our measurements. 
After establishing uniform DC conduction as a baseline, we analyze time-harmonic AC transport through a phenomenon-based approach without relying on advanced vector calculus.
By focusing on its two limiting cases, negligible current displacement and strong surface confinement, we mirror the physically intuitive treatment typically employed in introductory physics and engineering courses (e.g., \cite{Lueders,Grimsehl,Hayt}). 
The full analytical treatment is provided in the supplement (with further details in \cite{Simonyi,Landau,Kupfmuller}).

At the continuum level, the local behaviour of a homogeneous, linear conductor is governed by:
\begin{equation}
\vec{j}(\vec{r}, t) = \sigma \vec{E}(\vec{r}, t) \, , 
\label{eq:local_ohm}
\end{equation}
where $\vec{j}$ is the local current density, $\vec{E}$ the electric field at position $\vec{r}$ and time $t$, and $\sigma = 1/\rho$ the static electrical conductivity (with $\rho$ being the electrical resistivity).

Integrating the local fields in Equation~(\ref{eq:local_ohm}) over the conductor geometry relates the macroscopic voltage $U$ and current $I$. In the DC case, this yields the familiar macroscopic Ohm's law,
\begin{equation}
U = R_{\mathrm{DC}} I \, ,\quad \mathrm{with}  \quad R_{\mathrm{DC}} = \rho \frac{l}{A} 
\label{eq:ohm}
\end{equation}
as the static resistance of a homogeneous conductor of length $l$ and uniform cross-sectional area $A$.
For a solid cylindrical conductor of outer radius $r_0$, the conductive area is given by $A = \pi r_0^2$. 
In introductory physics courses, Equation~(\ref{eq:ohm}) is often introduced empirically.
Commercial setups similar to ours are widely used to measure small DC resistances and demonstrate how $R_{\mathrm{DC}}$ varies with geometry and material \cite{phywe}.
However, because they are traditionally equipped with non-magnetic specimens (e.g., copper or brass), they are generally not applied to investigate AC phenomena such as the skin effect.

Under AC excitation at frequency $f$ ($\omega = 2\pi f$), time-varying magnetic fields induce internal eddy currents that oppose current flow in the conductor core and reinforce it near the boundary. Consequently, $\vec{j}(\vec{r}, t)$ becomes spatially non-uniform, restricting current transport to a thin surface layer. Electrodynamically, this current redistribution and the resulting internal inductance transform the static resistance into a frequency-dependent complex internal impedance:
\begin{equation}
Z_{\mathrm{AC}} (\omega) = R_{\mathrm{AC}} (\omega) + i\, \omega L_{\mathrm{int}} (\omega) \, .
\label{eq:complex_impedance}
\end{equation}
Here, $R_{\mathrm{AC}} \geq R_{\mathrm{DC}}$ accounts for enhanced Joule dissipation in the reduced effective cross-section, and the reactive term $\omega L_{\mathrm{int}}$ reflects phase shifts due to internal magnetic flux, where $L_{\mathrm{int}}$ is the internal self-inductance. 
As the skin effect screens fields from the core, it simultaneously increases $R_{\mathrm{AC}}$  by lowering the effective conduction area and reduces $L_{\mathrm{int}}$ by decreasing the energy stored in the internal magnetic field.

The macroscopic relation between AC voltage and current thus reads
\begin{equation}
U = |Z_{\mathrm{AC}}| \, I \, ,
\label{eq:macroscopic_ac_ohm}
\end{equation}
allowing $|Z_{\mathrm{AC}}|$ to be extracted as the linear slope of $U(I)$ characteristics, analogous to the DC baseline. Here, $U$ and $I$ denote root-mean-square (RMS) amplitudes as recorded by standard multimeters.
Note that in our line-frequency setup intentionally varying the area of the voltage-pickup loop produces no discernible change in the voltage readout, confirming that external loop inductance $L_{\mathrm{ext}}$ is negligible.

Modeling the measured AC impedance $Z_{\mathrm{AC}}$ requires accounting for the spatial distribution of the internal fields. The characteristic length scale governing their spatial decay is the skin depth $\delta$, defined as 
\begin{equation}
\delta = \sqrt{\frac{2\rho}{\omega \mu_0 \mu_\mathrm{r}}} \, ,
\label{eq:skin_depth}
\end{equation}
where $\mu_0$ is the vacuum permeability and $\mu_\mathrm{r}$ is the relative magnetic permeability of the conductor. Physically, $\delta$ represents the depth at which both the current density and the driving electric field decay to $1/e$ ($\approx 37\%$) of their surface values, where $e$ is Euler's number.
For non-magnetic conductors, such as copper, $\delta$ typically ranges from about 10 mm at line frequency down to tens of micrometers in the MHz regime.

The electrodynamic response of a cylindrical conductor is governed by the dimensionless ratio of its physical radius $r_0$ to the skin depth $\delta$:
\begin{equation}
x = \frac{r_0}{\delta} \, .
\label{eq:x_parameter}
\end{equation}
In the quasistatic regime ($x \ll 1 \Rightarrow \delta \gg r_0$), 
the skin depth far exceeds the conductor dimensions. 
The current density remains essentially uniform across the conductor cross-section, yielding
\begin{equation}
\frac{R_{\mathrm{AC}}}{R_{\mathrm{DC}}} \approx 1 \, ,
\label{eq:lowf}
\end{equation}
while the internal self-inductance approaches its static DC value, $L_{\mathrm{int}} \approx \mu_0 \mu_\mathrm{r} l/(8\pi)$. However, the inductive reactance remains negligible relative to the ohmic resistance ($\omega L_{\mathrm{int}} \ll R_{\mathrm{AC}}$), implying that $|Z_{\mathrm{AC}}| \approx R_{\mathrm{AC}}$.
This explains why for standard copper wires with millimeter-scale diameters typically found in electrical installations, the skin effect is negligible at line frequency.

Conversely, in the limit of a strongly developed skin effect ($x \gg 1 \Rightarrow \delta \ll r_0$), current transport is confined to a thin surface boundary layer, resulting in the linear asymptote
\begin{equation}
\frac{R_{\mathrm{AC}}}{R_{\mathrm{DC}}} \approx \frac{\omega L_\mathrm{int}}{R_{\mathrm{DC}}} \approx \frac{x}{2} = \frac{r_0}{2\delta} \, .
\label{eq:high_freq_asymptote}
\end{equation}
Note that in this regime the inductive reactance is no longer negligible; as the internal self-inductance decreases with increasing frequency according to $L_{\mathrm{int}} = \mu_0 \mu_\mathrm{r} \, \delta \, l / (4\pi r_0) \propto 1/\sqrt{\omega}$, the inductive reactance asymptotically approaches the AC resistance, corresponding to a characteristic phase angle of $45^\circ$ and implying $|Z_{\mathrm{AC}}| \approx \sqrt{2}\, R_{\mathrm{AC}}$.

In the intermediate transition regime ($x \approx 1$, where $\delta \approx r_0$), the decaying current profiles from opposing boundaries overlap significantly in the core. In this region, the Bessel functions governing the field distribution cannot be reduced to simple algebraic approximations, requiring the full closed analytical expression provided in the supplement.

While our analysis relies on the rigorous analytical field theory for extended conductors, we note that such systems can alternatively be modeled via discrete ladder networks of concentric resistances and coupled inductances \cite{Wheeler,Kim}. This approach reduces the underlying field equations to a system of linear equations that approaches the analytical result in the continuum limit.

\section{Experimental section}
\label{sec:setup}

This section describes the physical properties of the investigated conductor specimens and the architecture of the electrical circuit for evaluating the resistance ratios at line frequency.

\subsection{Conductor materials}

As established by Equation~(\ref{eq:x_parameter}), observing pronounced skin effects at line frequency requires shifting the dimensionless parameter $x$ into the moderate or high regime ($x \geq 1$). Since copper features $\delta \approx 10\,\mathrm{mm}$, non-magnetic conductors would demand impractically thick, costly centimeter-scale rods (or even costlier silver, which has a lower resistivity).
Alternatively, $\delta$ can be drastically reduced via $\mu_\mathrm{r}$ (Equation~(\ref{eq:skin_depth})). While high-permeability alloys like Mu-metal are optimal in theory, they are expensive and mechanically sensitive.
Utilizing readily available, centimeter-thick laboratory support rods made of structural steel successfully scales the parameter $x$ into an accessible regime at line frequency, making them an ideal and cost-effective choice for demonstration experiments.
To highlight the role of magnetic permeability, comparing this behaviour with an identical geometry made of a non-magnetic material, such as stainless steel support rods, provides a striking experimental contrast.

The experimental investigation relies on cylindrical laboratory support rods ($\approx 100\,\mathrm{cm}$ length, nominal outer diameter $12\,\mathrm{mm}$) made from the two contrasting materials introduced above: non-magnetic austenitic stainless steel (AISI 304, commonly referred to as V2A) and ferromagnetic structural steel (S235JR, equivalent to ASTM A36). Outer dimensions were verified using a standard vernier caliper. The qualitative magnetic response was confirmed using an AlNiCo bar magnet, a simple demonstration where the magnet adheres firmly to the structural steel but exhibits no measurable attraction to the stainless steel. The geometric dimensions, nominal material parameters, and resulting electrodynamic quantities at $50\,\mathrm{Hz}$ are compiled in Table~\ref{tab:material_properties}.

\begin{table}[htbp]
\centering
\caption{Geometric, electrical, and magnetic properties of the tested laboratory conductors at line frequency ($f = 50\,\mathrm{Hz}$): measured specimen radius $r_0$, nominal specific electrical resistivity $\rho$ and relative magnetic permeability $\mu_\mathrm{r}$, estimated skin depth $\delta$ and electrodynamic parameter $x$. }
\label{tab:material_properties}
\begin{tabular}{l c c c c c}
\hline
Material & $r_0 \;/\; \mathrm{mm}$ &  $\rho \;/ \; \Omega\,\mathrm{mm}^2/\mathrm{m}$ & $\mu_\mathrm{r}$ & $\delta \;/\; \mathrm{mm}$ & $x$ \\
\hline
Stainless steel \cite{din} & 6.00(2)  & 0.73 & $\le 1.02$ & 64 & 0.1 \\
Structural steel \cite{asm} & 6.00(2)  & 0.16 & $\sim 10^3$ & 1 & 6 \\
\hline
\end{tabular}
\end{table}

The physical contrast of the conductor materials stems directly from their atomic crystal structures. In austenitic stainless steel, chromium ($\approx 18\%$) and nickel ($\approx 8\%$) preserve a face-centered cubic (fcc) lattice at room temperature. This structure suppresses parallel spin alignment ($\mu_\mathrm{r} \approx 1$) and enhances electron scattering via foreign-atom substitution, raising the electrical resistivity. At $50\,\mathrm{Hz}$, the resulting skin depth $\delta$ exceeds the rod radius $r_0$ by an order of magnitude, yielding a uniform current density distribution with no observable skin effect. Note that mechanical drawing can induce slight strain-induced martensitic transformations, causing localized permeability increases up to $\mu_\mathrm{r} \le 1.02$.
Conversely, structural steel retains the body-centered cubic (bcc) ferrite lattice of iron, exhibiting strong ferromagnetism ($\mu_\mathrm{r} \sim 10^3$) and significantly lower resistivity. The high magnetic permeability confines the electromagnetic field to a narrow surface layer, likely causing a pronounced skin effect at line frequency.

\subsection{Circuit layout}

The electrical setup is engineered around standard laboratory equipment to perform precise measurements on milliohm-scale conductors without requiring specialized high-frequency instrumentation. 
Excitation currents in the ampere regime are selected to yield potential drops in the millivolt range, securing a high signal-to-noise ratio easily resolvable by standard digital multimeters (Voltcraft VC 831; 0.1 mV resolution, input impedance $R_{\mathrm{vm}} = 10\,\mathrm{M}\Omega$).
Designed for line frequencies ($50\,\mathrm{Hz}$), these instruments provide accurate true-RMS readings without costly high-bandwidth devices.
Drive power is supplied by a variable AC/DC source ($25\,\mathrm{V}\,\mathrm{AC}$ / $20\,\mathrm{V}\,\mathrm{DC}$; e.g., PeakTech 6120). 
Driving currents up to $3\,\mathrm{A}$ through the rods yields a maximum surface magnetic flux density of roughly $0.1\,\mathrm{mT}$ for stainless steel compared to approximately $50\,\mathrm{mT}$ for structural steel, well below saturation induction, which typically lies in the tesla regime.
Electrical connections to the specimen are established via four heavy-duty crocodile clips wired with standard $4\,\mathrm{mm}$ laboratory banana leads. 
All measurements were performed at room temperature ($24\,^\circ\mathrm{C}$).

\begin{figure}[h]
  \centering
  \begin{minipage}[b]{0.48\textwidth}
    \centering
    \includegraphics[width=\textwidth]{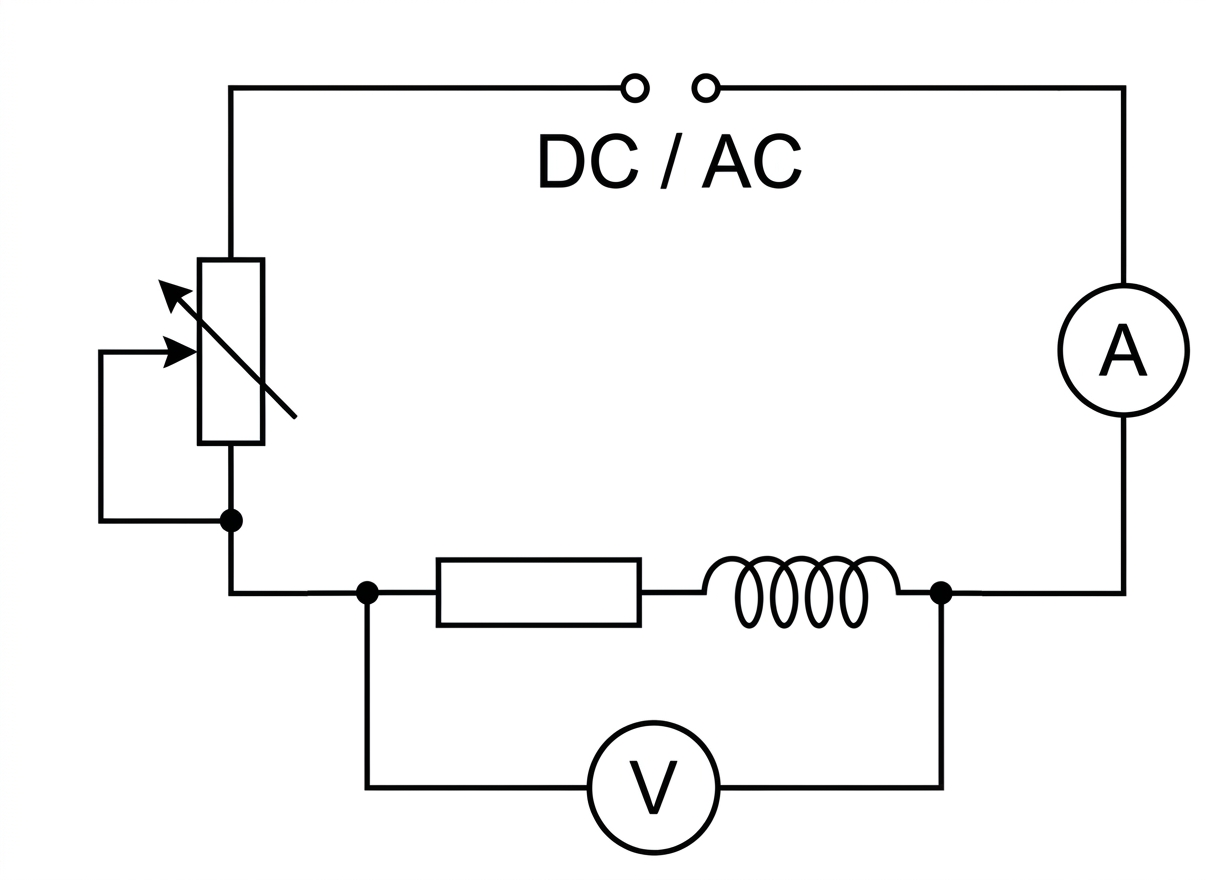}
    \centerline{(a)}
  \end{minipage}
  \hfill
  \begin{minipage}[b]{0.48\textwidth}
    \centering
    \includegraphics[width=\textwidth]{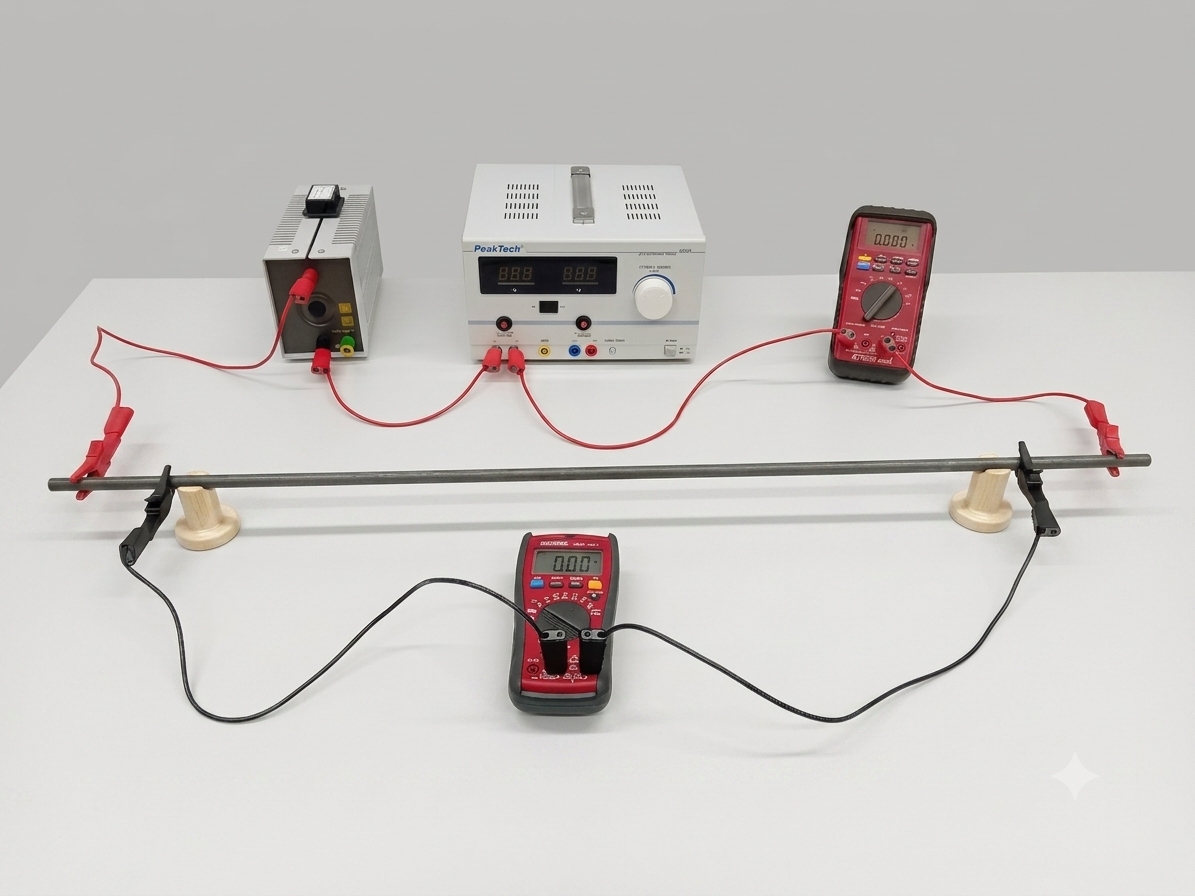}
    \centerline{(b)}
  \end{minipage}
\caption{(a) Schematic circuit diagram of the four-point probing setup (power source, series rheostat, ammeter A, and voltmeter V). The extended conductor specimen is represented by a lumped resistor in series with an internal inductance, accounting for the inductive reactance that arises under AC operation. (b) Practical laboratory implementation: the outer red crocodile clips inject the driving current from the variable source via the rheostat and the ammeter, while the inner black crocodile clips tap the potential drop across a defined gauge length directly into the voltmeter.}
  \label{fig:setup}
\end{figure}

The schematic layout and laboratory implementation are depicted in Figure~\ref{fig:setup}(a) and (b), respectively. 
In the equivalent circuit, the conductor specimen is represented by a lumped resistor in series with an internal inductance. Under steady-state DC conditions, this inductance generates no voltage drop and does not affect the $R_{\mathrm{DC}}$ measurement. Under AC operation, however, the internal inductive reactance can reach a magnitude comparable to the AC resistance, requiring consideration when modeling the overall electrical response.
Connecting a voltage source directly across a milliohm specimen would create a near-short-circuit condition. To stabilize the current and prevent rapid Joule heating, a high-power slide rheostat  ($10\,\Omega$, $5.7\,\mathrm{A}$ continuous rating; Indosaw PRN-322 / Contrex) is placed in series as a ballast resistor, converting the supply into a passive constant current source. 
Primary current scaling is controlled via the power supply, with the rheostat pre-set while de-energized to avoid thermal hysteresis or contact arcing. 
To eliminate parasitic contact resistances, a four-point (Kelvin) configuration is employed. 
Two outer red crocodile clips for current injection are attached with a clearance of at least $6\,\mathrm{cm} \gg r_0$ from the voltage taps to ensure uniform current density. 
Completely decoupled from the drive loop, two inner black crocodile clips tap the potential drop across a defined gauge length $l = 80.0\,\mathrm{cm}$ (measured between their inner edges using a class~II tape measure). 
Full $U(I)$ characteristics (15 setpoints up to $3\,\mathrm{A}$) are recorded under both DC and $50\,\mathrm{Hz}$ AC excitation, taking prompt readings to prevent ohmic self-heating and thermal drift.

\section{Results and discussion}
\label{sec:results}

This section first establishes the static DC properties of the solid cylindrical conductors and then analyzes the material-dependent AC impedance surge at line frequency ($50\,\mathrm{Hz}$), highlighting the role of permeability in driving the skin effect.

\subsection{DC baseline and material resistivity}

Prior to investigating AC transport, a DC baseline measurement was established to verify Ohm's law and determine the static electrical properties of the conductors (Equation (\ref{eq:ohm})). 
Measurements were performed on solid rods of non-magnetic stainless steel and ferromagnetic structural steel with a conductive length $l = 80.0\,\mathrm{cm}$ and rod radius $r_0 = 6.00\,\mathrm{cm}$.

\begin{figure}[h]
  \centering
  \includegraphics[width=0.65\textwidth]{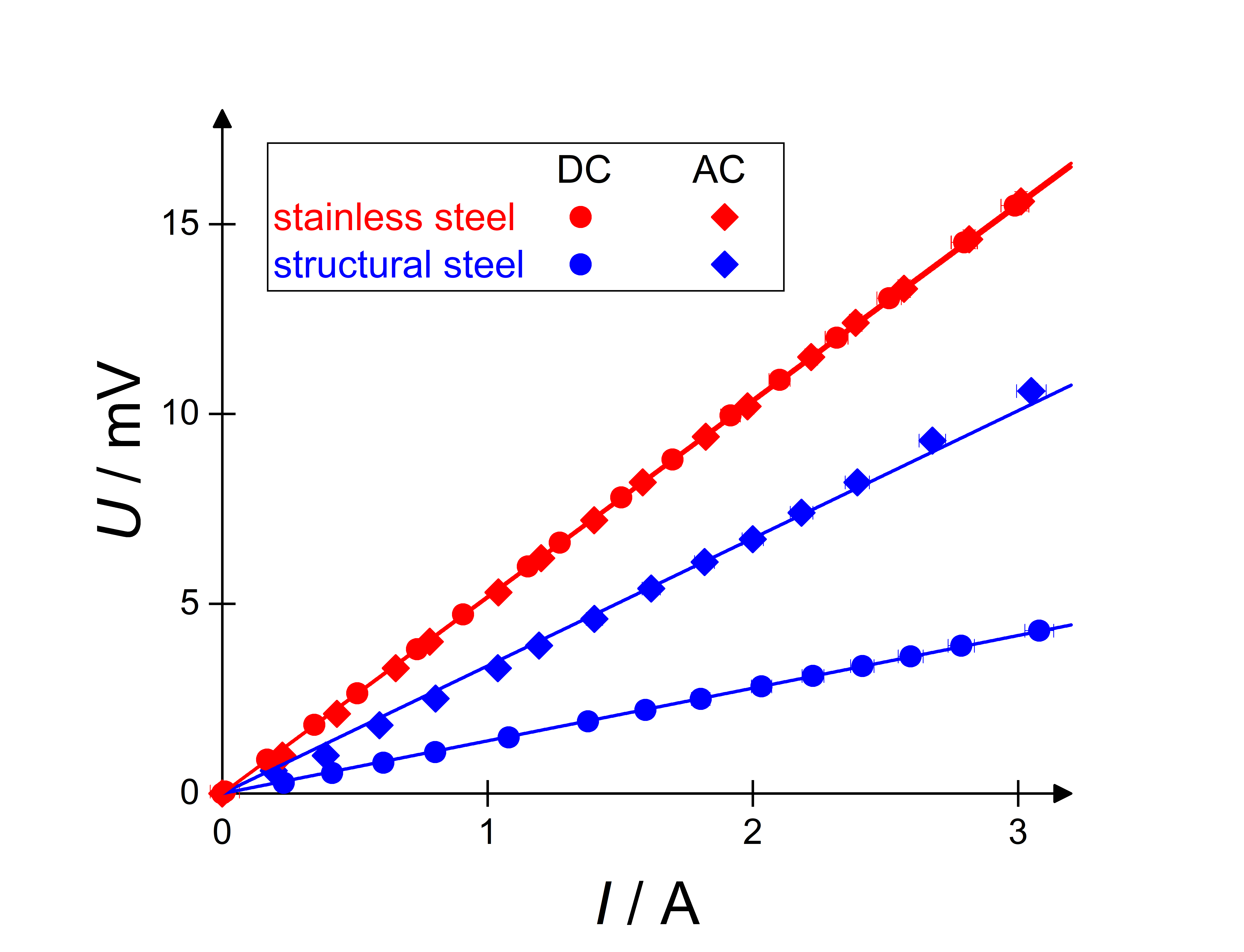}
  \caption{Voltage--current ($U$--$I$) characteristics for solid cylindrical rods ($l = 80.0\,\mathrm{cm}$, $r_0 = 6.00\,\mathrm{mm}$) under DC (circles) and $50\,\mathrm{Hz}$ AC excitation (diamonds). 
  For non-magnetic stainless steel (red), DC and AC characteristics collapse onto a single line ($|Z_{\mathrm{AC}}| \approx R_{\mathrm{DC}}$). 
  For ferromagnetic structural steel (blue), high permeability induces a strong skin effect, causing a sharp surge in the AC slope ($|Z_{\mathrm{AC}}| > R_{\mathrm{DC}}$). Solid lines represent single-parameter linear fits through the origin; uncertainties are smaller than the symbol size where not visible.}
  \label{fig:results}
\end{figure}

As shown by the circular symbols in Figure~\ref{fig:results}, the DC $U$--$I$ characteristics exhibit strictly linear behaviour passing through the origin. The slope directly yields the static DC resistance $R_{\mathrm{DC}}$ (Equation (\ref{eq:ohm})), obtained via linear fits with a fixed zero intercept. Stainless steel displays a significantly steeper DC slope ($R_{\mathrm{DC}} = 5.19(1)\,\mathrm{m}\Omega$) than structural steel ($R_{\mathrm{DC}} = 1.39(1)\,\mathrm{m}\Omega$).

Based on Equation~(\ref{eq:ohm}) and the conductor geometry, the static resistivities were extracted: $\rho = 0.734(5)\,\Omega\,\mathrm{mm}^2/\mathrm{m}$ for stainless steel and $\rho = 0.196(2)\,\Omega\,\mathrm{mm}^2/\mathrm{m}$ for structural steel (Table~\ref{tab:dc}). 
Both values agree with literature references within experimental uncertainty, with stainless steel matching nominal expectations precisely and structural steel exhibiting a minor, batch-dependent deviation (Table~\ref{tab:material_properties}).

The high accuracy of these DC results, as reflected in the agreement with literature data, relies directly on four-point Kelvin sensing, which suppresses parasitic lead and contact resistances.
While commercial high-precision educational setups \cite{phywe} achieve submilliohm resolution using sensitive signal amplifiers, our approach generates directly readable millivolt signals on standard laboratory multimeters simply by employing high currents and large conductor specimens. This eliminates the need for specialized amplification electronics while using readily available, low-cost structural materials, rendering precise four-point metrology accessible for large student cohorts.

\begin{table}[h]
\caption{Empirical DC resistance extracted from linear fits to the DC data (circles in Figure \ref{fig:results}) and resulting specific electric resistivities $\rho$ for conductor samples.}
\centering
\begin{tabular}{l l c c }
\hline
Material & Geometry &  $R_{\mathrm{DC}}$ / m$\Omega$ & $\rho \;/ \; \Omega\,\mathrm{mm}^2/\mathrm{m}$  \\
\hline
Stainless steel & Solid rod     & $5.19(1)$  & $0.734(5)$  \\
Structural steel & Solid rod        & $1.39(1)$  & $0.196(2)$   \\
\hline
\end{tabular}
\label{tab:dc}
\end{table}

\begin{table}[h]
\caption{Experimental DC resistance ($R_{\mathrm{DC}}$) and AC impedance ($|Z_{\mathrm{AC}}|$), both extracted from linear fits (Fig.~\ref{fig:results}), along with AC resistance and impedance ratios derived from the asymptotic Equations (\ref{eq:lowf}) and (\ref{eq:high_freq_asymptote}).}
\centering
\begin{tabular}{l  c c c c}
\hline
Material  &  $R_{\mathrm{DC}}$ / m$\Omega$ & $|Z_{\mathrm{AC}}|$ / m$\Omega$ & $|Z_{\mathrm{AC}}| / R_{\mathrm{DC}}$  &  $R_{\mathrm{AC}} / R_{\mathrm{DC}}$  \\
\hline
Stainless steel & $5.19(1)$  & $5.16(2)$ & $0.998(5)$ &  $0.998(5)$    \\
Structural steel &  $1.39(1)$  & $3.36(4)$ & $2.42(3)$  &$1.71(2)$ \\
\hline
\end{tabular}
\label{tab:acdc}
\end{table}

\subsection{Material contrast under AC conditions: impact of magnetic permeability on the skin effect}

The AC experiments (diamond symbols in Figure~\ref{fig:results}) compare the electrical response of the solid rods under $50\,\mathrm{Hz}$ AC excitation. 
While the AC characteristics remain linear through the origin, their slope reflects the magnitude of the complex AC impedance $|Z_{\mathrm{AC}}|$ rather than pure static resistance (Equation (\ref{eq:macroscopic_ac_ohm})), with results listed in Table \ref{tab:acdc}.

For non-magnetic stainless steel ($\mu_{\mathrm{r}} \approx 1$), the AC and DC characteristics collapse onto a single line within experimental uncertainty ($|Z_{\mathrm{AC}}| / R_{\mathrm{DC}} = 0.998(5)$).  
At $50\,\mathrm{Hz}$, the calculated skin depth ($\delta \approx 64\,\mathrm{mm}$) vastly exceeds the rod radius ($r_0 = 6.00\,\mathrm{mm}$). 
Consequently, current density remains uniform throughout the cross-section. 
In this quasistatic regime ($x \ll 1$), the inductive reactance is negligible compared to the ohmic resistance, yielding $|Z_{\mathrm{AC}}| \approx R_{\mathrm{AC}} \approx R_{\mathrm{DC}}$, as expected based on Equation (\ref{eq:lowf}). 
In terms of electrical transport, the conductor thus behaves identically under AC and DC conditions.

In sharp contrast, ferromagnetic structural steel ($\mu_{\mathrm{r}} \gg 1$) shows a clear increase in slope under AC conditions.
Although structural steel is the superior DC conductor 
(with a static resistivity lower than that of stainless steel by about a factor of 4, see Table~\ref{tab:dc}), its AC impedance magnitude rises to $|Z_{\mathrm{AC}}| = 3.36(4)\,\mathrm{m}\Omega$, exceeding its static resistance by a factor of $2.42(3)$. 
This striking inversion creates a cognitive conflict for students: despite stainless steel retaining a higher absolute resistance overall, structural steel experiences a distinct impedance increase at $50\,\mathrm{Hz}$, whereas stainless steel remains unaffected.
Physically, high magnetic permeability compresses the skin depth to the millimeter scale ($\delta < r_0$), placing the system well into the skin-effect regime.
Consequently, structural steel reaches pronounced skin-effect behaviour at a mere $50\,\mathrm{Hz}$, whereas non-magnetic conductors such as stainless steel, copper, or aluminum would require frequencies in the kilohertz or megahertz range to exhibit a comparable electrodynamic response.

The behaviour is well-approximated by the high-frequency expansion, where two distinct electrodynamic mechanisms contribute to the observed impedance surge, according to Equation (\ref{eq:high_freq_asymptote}). 
First, the ohmic resistance increases because current is confined to a thin surface layer of effective thickness $\delta$, reducing the conducting cross-section. 
Second, the magnetic field trapped within the conductor core induces a substantial internal inductive reactance. 
Under these conditions, the internal reactance closely approaches the ohmic resistance, yielding an overall impedance magnitude of $|Z_{\mathrm{AC}}| \approx \sqrt{2} \, R_{\mathrm{AC}}$.
Within this strong skin-effect limit, the ohmic component can be isolated according to $R_{\mathrm{AC}} \approx |Z_{\mathrm{AC}}| / \sqrt{2} = 2.38(3)\,\mathrm{m}\Omega$, yielding an ohmic resistance ratio of $R_{\mathrm{AC}}/R_{\mathrm{DC}} = 1.71(2)$.

While the observed impedance surge can be heuristically understood based on the strong skin-effect formulas, the full analytical solution is considered for a rigorous characterization of the material response. 
Utilizing Equation (S8) from the supplement with the measured resistivity $\rho$ and treating $\mu_\mathrm{r}$ as an effective free parameter, the experimental impedance ratio is precisely recovered.
This procedure yields $R_{\mathrm{AC}} / R_{\mathrm{DC}} \approx 1.83$ (thus confirming the asymptotic trend estimated above), along with $\omega L_{\mathrm{int}} / R_{\mathrm{DC}} \approx 1.56$, $\delta \approx 1.5\,\mathrm{mm}$ and $\mu_{\mathrm{r}} \approx 450$.

Note that for ferromagnetics, $\mu_{\mathrm{r}}$ is not a static material constant, but an effective, operating-point value dependent on the applied current $I$ and the resulting internal magnetic field strength.
The experimentally derived effective permeability is significantly lower than typical DC literature nominals ($\mu_{\mathrm{r}} \sim 10^3$). 
This discrepancy can be primarily attributed to nonlinear hysteresis and skin-effect-induced eddy currents under AC conditions.

A closer inspection of the linear fit to the AC $U(I)$ characteristics of structural steel reveals small, systematic deviations: data points at lower currents ($I \leq 1.4\,\mathrm{A}$) lie slightly below the fit, whereas those at higher currents ($I \geq 2.4\,\mathrm{A}$) trend slightly above it. This behaviour reflects the current-dependent nature of the effective permeability in the un-saturated regime. As the current amplitude increases, the progressive alignment of magnetic domains enhances the magnetic response, causing the local slope (and thus the AC impedance) to increase. While core losses and hysteresis effects also play a role at elevated field strengths, this current-dependent stiffening of the effective permeability accounts for the observed slight upward curvature.

\section{Pedagogical design and implementation}
\label{sec:didactics}

To maximize conceptual learning and engagement, the design and deployment of the experiment follow established pedagogical principles. This section synthesizes the core educational insights, outlines the visual strategies used to minimize cognitive load, details the learning sequence for cognitive activation, and presents its implementation as a fast-paced demonstration.

\subsection{Educational synthesis}

The presented experiment offers clear pedagogical benefits by transforming the skin effect from an abstract mathematical concept into an intuitive physical phenomenon.
In standard curricula, the skin effect is often incorrectly pigeonholed as a high-frequency phenomenon irrelevant at line frequency. By using structural steel and a diameter of 12 mm, this setup directly challenges that misconception, demonstrating that a high relative permeability shifts the characteristic frequency scale down to $50\,\mathrm{Hz}$.
Furthermore, the experiment addresses student cognitive conflicts regarding material performance: learners frequently expect the superior DC conductor (structural steel) to outperform stainless steel under all conditions, only to observe its AC impedance surge at line frequency due to high permeability.

\subsection{Visual design and cognitive load reduction}

The physical setup of a demonstration experiment influences how effectively learners process physical concepts. 
According to cognitive load theory, instructional design must minimize extraneous cognitive load to avoid overwhelming the working memory during conceptual processing \cite{Paas}.
In a demonstration, clear spatial structuring enhances perceptual accessibility \cite{Schmidkunz}, whereas extraneous elements act as seductive details that hinder understanding \cite{Sundararajan}. 
Optimizing a lecture demonstration thus requires highlighting essential physical features while eliminating unnecessary components.

Our experimental setup (Figure~\ref{fig:setup}) directly translates schematic topology into an intuitive spatial arrangement. 
Mounting the sample rod horizontally across the center provides a clear visual focus, with core functional components aligned along the central vertical axis according to their visual depth.
The voltmeter occupies the central foreground, while the power supply is positioned in the background plane. 
Secondary control elements, such as the rheostat and ammeter, are shifted to the rear left and right peripheries. 
This visual hierarchy directs attention toward the primary observables while mirroring the theoretical circuit diagram to eliminate split-attention effects. 
Furthermore, wiring is strictly colour-coded and spatially segmented.
High-current supply cables run along the rear plane, attaching to the outer ends of the conductor, while potential measuring leads occupy the front plane, tapping symmetrically along the conductor.
This structure enables students to trace current and potential paths effortlessly without visual clutter. 
Finally, replacing specialized high-frequency equipment with standard laboratory instruments in a four-point setup maintains structural transparency, allowing learners to focus on the physical phenomenon rather than navigating instrument interfaces.

\subsection{Cognitive activation via Predict--Observe--Explain strategies}

Simply performing a physics demonstration does not guarantee learning; passive observation is often ineffective, leaving conceptual understanding no better than omitting the experiment entirely \cite{Champagne, Crouch}. 
Learning gains depend critically on how a demonstration is integrated into the lecture. To foster genuine conceptual change, instructional design should actively engage learners through a Predict--Observe--Explain (POE) sequence \cite{ Sokoloff,White}. 
The quantitative contrast between materials in our setup provides an ideal framework for such cognitive activation. 
When deployed as an interactive lecture demonstration, the setup serves as a powerful catalyst for cognitive activation. 
Before the measurement, the instructor prompts students to predict the relative AC impedance of both materials compared to their DC baseline (greater, smaller, or equal), requiring them to commit to a choice via live polling or immediate written notes. The subsequent measurement confronts them with a dramatic, counterintuitive result: the statically superior conductor (structural steel) exhibits a sharp surge in AC voltage drop, whereas the stainless steel rod remains unchanged. 
Because this discrepancy cannot be resolved using familiar static DC models, it induces a cognitive conflict, forcing learners to re-evaluate their assumptions regarding electric transport and material properties. 
This conceptual conflict is subsequently resolved by introducing the skin depth $\delta$ as a permeability-dependent length scale.

\subsection{Implementation as a fast-paced demonstration}

When adapted for a quick lecture demonstration, the procedure is streamlined for maximum visual impact and execution speed. Large-scale analog demonstration meters (or projected digital displays) replace standard multimeters, and the full $U(I)$ sweep is reduced to a single-point comparison at a fixed reference current (e.g., $I = 3.0\,\mathrm{A}$, aligned with full-scale ammeter deflection). This reduces cognitive load, allowing students to focus entirely on the voltmeter reading. 
To fit even the tightest lecture schedules, the experiment can be completed in under two minutes by showcasing the dramatic AC impedance surge on a structural steel sample alone -- or, if time permits, by contrasting its response directly with non-magnetic stainless steel.

\section{Conclusion}
\label{sec:conclusion}

We have presented a simple, cost-effective experimental setup that brings the skin effect into line-frequency undergraduate laboratories using standard equipment and milliohm-scale four-point Kelvin sensing. 
The quantitative measurements confirm a key electrodynamic insight: 
High magnetic permeability drastically compresses the skin depth $\delta$ at just $50\,\mathrm{Hz}$, inducing a dramatic surge in AC impedance for structural steel, whereas non-magnetic stainless steel remains unaffected. 
By combining this unexpected material behaviour with the realization that the skin effect is not restricted to high-frequency domains, the experiment challenges common student misconceptions about AC electrical transport. Without requiring specialized high-frequency apparatus, this approach provides a versatile, intuitive tool for both interactive lecture demonstrations and quantitative laboratory experiments.

\ack{We thank R.~Platten for assistance with the experiments, L.~Keminer for critical reading of the manuscript, and the mechanical workshop for providing metal rods.}

\funding{This work benefited from funding through the ZSL program.}

\roles{H.W. originated the concept for this study. F.P. expanded the scope, designed and performed the experiments, analyzed the data, and wrote the initial manuscript. Both authors reviewed and approved the final text.}

\data{All data supporting the findings of this study are included within the article and its supplementary material.}

\suppdata{Supplementary material includes a brief theoretical derivation of the AC impedance of cylindrical conductors.}

\bibliographystyle{iopart-num}

\newpage

\setcounter{page}{1}
\setcounter{equation}{0}
\setcounter{table}{0}
\setcounter{figure}{0}
\setcounter{section}{0}

\renewcommand{\thepage}{S\arabic{page}}
\renewcommand{\theequation}{S\arabic{equation}}
\renewcommand{\thefigure}{S\arabic{figure}}
\renewcommand{\thetable}{S\arabic{table}}
\renewcommand{\thesection}{S\arabic{section}}

\title{Why the better conductor surges in impedance:\\
a low-cost demonstration of the skin effect at line frequency
\\
-- Supplementary material --}

\author{Florian Platten$^{1,*}$ and Helmut Wenz$^{1}$}

\affil{$^1$Faculty of Mathematics and Natural Sciences, Heinrich Heine University D\"{u}sseldorf, D\"{u}sseldorf, Germany}

\affil{$^*$Author to whom any correspondence should be addressed.}

\email{florian.platten@hhu.de}

\noindent\rule{\linewidth}{0.1pt}
\vspace{.5cm}

\section{Theoretical details}
\subsection{Derivation of the field equation in cylindrical conductors}

At the continuum level, the local behaviour of a homogeneous, linear conductor is governed by 
\begin{equation}
\vec{j}(\vec{r}, t) = \sigma \vec{E}(\vec{r}, t) \,  ,
\label{eq:si_locohm}
\end{equation}
where homogeneity assumes a spatially uniform electrical conductivity $\sigma$, and linearity ensures that $\sigma$ remains independent of the electric field strength $\vec{E}$.
Deviations from this local approximation only become relevant under extreme physical regimes, such as high-field thermal nonlinearities, high-frequency kinetic effects exceeding the electron relaxation time $\tau$ ($\omega \tau \gg 1$), or cryogenic conditions where the mean free path exceeds the skin depth (anomalous skin effect) -- none of which apply to the present experiment.
Equation~(\ref{eq:si_locohm}) holds locally for both DC and the quasistationary AC regime, as electron relaxation times in metals are orders of magnitude faster than the field period.

Technical metals typically feature exceptionally high electrical conductivity.
Even at megahertz frequencies, the displacement current density 
$\vec{j}_{\mathrm{d}}  = \varepsilon_0 \varepsilon_\mathrm{r} \partial \vec{E}/\partial t$ 
(with $\varepsilon_0$ being the vacuum permittivity and $\varepsilon_{\mathrm{r}}$ 
the relative permittivity)
is negligible compared to the conduction current density 
(Equation (\ref{eq:si_locohm})):
\begin{equation}
\left|\frac{ \vec{j}_{\mathrm{d}} }{ \vec{j} } \right| =  \frac{\omega  \varepsilon_0 \varepsilon_\mathrm{r}}{\sigma} \ll 1 \, .
\end{equation}
Under this quasistationary approximation, Maxwell's equations 
for the electric field \(\vec{E}\) and the magnetic flux density \(\vec{B}\) in a linear, homogeneous, charge-free conductor reduce to:
\begin{equation}
\nabla \times \vec{E} = -\frac{\partial \vec{B}}{\partial t}\, , 
\quad \nabla \times \vec{B} = \mu_0 \mu_\mathrm{r} \sigma \vec{E} \, , \quad 
\nabla \cdot \vec{E} = 0 \, , 
\quad \nabla \cdot \vec{B} = 0 \, .
\end{equation}

Taking the curl of Faraday's law and applying $\nabla \times (\nabla \times \vec{E}) = \nabla(\nabla \cdot \vec{E}) - \nabla^2 \vec{E}$ yields the governing field equation inside the conductor:
\begin{equation}
\nabla^2 \vec{E} = \mu_0 \mu_\mathrm{r} \sigma \frac{\partial \vec{E}}{\partial t} \, .
\label{eq:diffusion_E}
\end{equation}
Equation~(\ref{eq:diffusion_E}) is a parabolic diffusion equation rather than a hyperbolic wave equation (which contains a second-order time derivative, $\partial^2 \vec{E}/\partial t^2$).
This mathematical transition directly reflects the difference in how electromagnetic energy interacts with matter. In non-conducting dielectrics, field propagation is strictly hyperbolic: an unattenuated wave periodically polarizes and magnetizes the surrounding molecules, causing the induced oscillating dipoles to emit secondary fields that combine with the primary wave to yield propagation at a reduced phase velocity without energy dissipation. In a good conductor, however, the presence of free charge carriers generates large conduction currents (Equation (\ref{eq:si_locohm})). These currents rapidly convert the field energy into heat via ohmic losses, preventing the formation of a self-sustaining wave. Consequently, the electromagnetic field does not propagate as a wave, but instead diffuses into the conductor while undergoing strong exponential attenuation, a mechanism mathematically identical to thermal conduction governed by Fourier's law.

For a time-harmonic excitation $\vec{E}(\vec{r}, t) = \mathrm{Re}\{ \vec{E}(\vec{r}) e^{i\omega t} \}$ in an infinitely long cylindrical conductor aligned along the $z$-axis, symmetry dictates that $\vec{E}$ has only an axial component $E_z(r)$ depending solely on the radial coordinate $r$. In cylindrical coordinates, Equation~(\ref{eq:diffusion_E}) simplifies to:
\begin{equation}
\frac{d^2 E_z(r)}{dr^2} + \frac{1}{r}\frac{d E_z(r)}{dr} - k^2 E_z(r) = 0 \, ,
\quad \mathrm{with}\,\, k = \sqrt{i \omega \mu_0 \mu_\mathrm{r} \sigma} = \frac{1+i}{\delta} \, .
\label{eq:bessel_dgl_E}
\end{equation}
Requiring $E_z(r)$ to remain finite at the conductor axis ($r = 0$) and matching the surface boundary condition $E_z(r_0) = E_0$ yields the spatial field distribution:
\begin{equation}
E_z(r) = E_0 \frac{J_0(k r)}{J_0(k r_0)} \, ,
\label{eq:electric_field_distribution}
\end{equation}
where $J_0$ is the zeroth-order Bessel function of the first kind.

\subsection{AC impedance of cylindrical conductors}

To transition from the local field distribution to measurable macroscopic quantities, the total alternating current $I$ is obtained by integrating the axial current density over the circular cross-section. 
Applying the recurrence relation $\int z J_0(z) \, \mathrm{d}z = z J_1(z)$ for the complex argument $z$, where $J_1$ denotes the first-order Bessel function of the first kind, yields:
\begin{equation}
I = \frac{2\pi \sigma r_0 E_0}{k} \frac{J_1(k r_0)}{J_0(k r_0)} \, .
\end{equation}
Considering a cylindrical conductor of length $l$, the complex internal AC impedance $Z_{\mathrm{AC}}$ is defined as the ratio of the surface voltage drop $U = E_z(r_0) \, l = E_0 \, l$ to the total current $I$:
\begin{equation}
Z_{\mathrm{AC}} = \frac{k\,l}{2\pi \sigma r_0} \frac{J_0(k r_0)}{J_1(k r_0)} = R_\mathrm{AC} + i \, \omega L_\mathrm{int} \, .
\label{eq:exact_Z_ratio}
\end{equation}
Normalizing $Z_{\mathrm{AC}}$ by the static DC resistance $R_{\mathrm{DC}} = l/(\pi r_0^2 \sigma)$ causes the conductor length $l$ to cancel out, yielding the dimensionless ratios for the AC resistance and internal reactance, respectively:\begin{equation}
\frac{R_{\mathrm{AC}}}{R_{\mathrm{DC}}} = \mathrm{Re}\left\{ \frac{k r_0}{2} \frac{J_0(k r_0)}{J_1(k r_0)} \right\}  \, ,
\end{equation}
\begin{equation}
\frac{\omega L_{\mathrm{int}}}{R_{\mathrm{DC}}} = \mathrm{Im}\left\{ \frac{k r_0}{2} \frac{J_0(k r_0)}{J_1(k r_0)} \right\}  \, .
\end{equation}
In 1D lumped-circuit theory, resistance $R$ and self-inductance $L$ are assumed to be fixed geometric constants. 
For 3D extended conductors, however, spatial field redistributions invalidate this abstraction: alternating fields induce internal eddy currents that concentrate current density near the boundary ($R_{\mathrm{AC}}$ increases) and expel the internal magnetic field ($L_{\mathrm{int}}$ decreases). 
As a result, the internal impedance becomes intrinsically frequency-dependent.
Remarkably, this entire frequency dependence is governed by $k r_0 = (1 + \mathrm{i}) \, x$. 
Consequently, all macroscopic dimensionless ratios depend solely on this single parameter $x$.

In the quasistatic regime ($x \ll 1$), the ratio of Bessel functions is expanded via Taylor series, yielding the low-frequency corrections:
\begin{equation}
\frac{R_{\mathrm{AC}}}{R_{\mathrm{DC}}} \approx 1 + \frac{1}{48} x^4  \, ,
\label{eq:low_freq_series}
\end{equation}
\begin{equation}
\frac{\omega L_{\mathrm{int}}}{R_{\mathrm{DC}}}  \approx 4 x^2  \, .
\label{eq:llow_freq_series}
\end{equation}
The constant leading-order term in $R_{\mathrm{AC}}/R_{\mathrm{DC}}$ agrees with Equation (8) in the main text.
At low frequencies, the first frequency-dependent correction to DC resistance scales quadratically with frequency ($\omega^2 \propto 1/\delta^4 \propto x^4$), reflecting very weak eddy-current opposition near the conductor center.

In the strong skin-effect regime ($x  \gg 1$), the ratio of Bessel functions 
can be expanded asymptotically, yielding the high-frequency expressions:
\begin{equation}
\frac{R_{\mathrm{AC}}}{R_{\mathrm{DC}}}  \approx \frac{x}{2} + \frac{1}{4} + \frac{3}{32 x}  \, ,
\label{eq:high_freq_series}
\end{equation}
\begin{equation}
\frac{\omega L_{\mathrm{int}}}{R_{\mathrm{DC}}} \approx \frac{x}{2} - \frac{3}{32 x} + \frac{3}{32 x^2}  \, .
\end{equation}
The leading linear term in $R_{\mathrm{AC}}/R_{\mathrm{DC}}$ corresponds to the planar surface limit and agrees with Equation (9) in the main text.

\end{document}